\documentclass[a4paper,fleqn,usenatbib]{mnras}

\usepackage{graphics}
\usepackage{graphicx}
\usepackage{psfig}
\input{epsf}
\usepackage{amsmath}
\usepackage{amssymb}
\usepackage{mathtools}
\usepackage{url}
\usepackage[T1]{fontenc}
\usepackage{ae,aecompl}
\usepackage{widetext}

\DeclarePairedDelimiter{\evdel}{\langle}{\rangle}
\newcommand{\ev}{\operatorname{}\evdel}
\newcommand{\be}{\begin{equation}}
\newcommand{\ee}{\end{equation}}

\newcommand{\order}[1]{\mathcal{O}\!\left(#1\right)}
\newcommand{\uS}{\mathrm{s}}

\title[Cosmic String Detection with Tree-Based Machine Learning]{Cosmic String Detection with Tree-Based Machine Learning}

\author[Vafaei et al.]{
A. Vafaei Sadr$^{1,2,3}$, M. Farhang$^{1}$, S. M. S. Movahed$^{1,4}$ \thanks{E-mail: m.s.movahed@ipm.ir}, B. Bassett$^{3,5,6,7}$, M. Kunz$^{2}$\\
$^1$ Department of Physics, Shahid Beheshti University, Velenjak, Tehran 19839, Iran\\
$^2$ D\'epartement de Physique Th\'eorique and Center for Astroparticle Physics, Universit\'e de Gen\`eve, 24 Quai Ernest Ansermet,\\ 1211 Gen\`eve 4, Switzerland\\
$^3$ African Institute for Mathematical Sciences, 6 Melrose Road, Muizenberg, 7945, South Africa\\
$^4$ School of Physics, Institute for Research in Fundamental Sciences, (IPM), P. O. Box 19395-5531, Tehran, Iran\\
$^5$ South African Astronomical Observatory, Observatory, Cape Town, 7925, South Africa\\ $^6$ SKA South Africa, The Park, Park Road, Pinelands, Cape Town 7405, South Africa\\
$^7$ Department of Maths and Applied Maths, University of Cape Town, Cape Town, South Africa}

\begin{document}
\label{firstpage}
\pagerange{\pageref{firstpage}--\pageref{lastpage}}
\maketitle

\begin{abstract}
We explore the use of random forest and gradient boosting, two powerful tree-based machine learning algorithms, for the detection of cosmic strings in maps of the cosmic microwave background (CMB), through their unique Gott-Kaiser-Stebbins effect on the temperature anisotropies.The information in the maps is compressed
into feature vectors before being passed to the  learning units. The feature vectors contain various statistical measures of  processed CMB maps that boost the cosmic string detectability. Our proposed classifiers, after training, give results improved over or similar to the claimed detectability levels of the existing methods for string tension, $G\mu$.
They can make $3\sigma$ detection of strings with $G\mu \gtrsim 2.1\times 10^{-10}$ for noise-free, $0.9'$-resolution CMB observations. The minimum detectable tension increases to $G\mu \gtrsim 3.0\times 10^{-8}$ for a more realistic, CMB S4-like (II) strategy, still a significant improvement over the previous results.
\end{abstract}



\section{INTRODUCTION}\label{intro}
The inflationary paradigm is the most widely accepted scenario for seeding the structures in the Universe, so far passing observational tests with flying colors. There is, however, both theoretical and observational room for contributions from alternative, well-motivated scenarios. Among these are perturbations sourced by cosmic topological defects formed at cosmological phase transitions.
 In particular, cosmic strings (CS) are theoretically expected to be produced in the early Universe \citep{Kibble:1976sj,Zeldovich:1980gh,Vilenkin:1981iu,
Vachaspati:1984dz,Vilenkin:1984ib,Shellard:1987bv,
Hindmarsh:1994re,Vilenkin:2000jqa,Sakellariadou:2006qs,
Bevis:2007gh,Depies:2009im,Bevis:2010gj,
Copeland:1994vg,Sakellariadou:1997zt,Sarangi:2002yt,
Copeland:2003bj,Pogosian:2003mz,Majumdar:2002hy,
Dvali:2003zj,Kibble:2004hq,HenryTye:2006uv}. 
The detection of  CS  would  open a unique window to the physics of the early Universe \citep{Kibble:1976sj,Zeldovich:1980gh,Vilenkin:1981iu,Vilenkin:2000jqa,Firouzjahi:2005dh}. Therefore a lot of effort has been put into developing powerful statistical tools for cosmic string network detection and  putting tight upper bounds on the CS tension, parametrized by $G\mu$, where $G$ and $\mu$ represent Newton's constant and the string's tension, respectively. 
The string tension is intimately related to the energy of the phase transition epoch,
\begin{equation}
\frac{G\mu}{c^2}=\order{\frac{\varpi^2}{M_{\rm Planck}^2}},
\end{equation}
where $\varpi$ is the symmetry breaking energy scale, $c$ is the speed of light and $M_{\rm Planck}\equiv\sqrt{\hbar c/G}$ represents the Planck mass. In this paper we work in natural units with $\hbar=c=1$.

A CS network would leave various imprints on cosmic microwave background (CMB) anisotropies. The Gott-Kaiser-Stebbins (KS) effect \citep{Kaiser:1984iv,Gott:1985, Stebbins:1988, Bouchet:1988hh, Allen:1997ag,Pen:1997ae,Ringeval:2012tk} corresponds to the integrated Sachs-Wolfe effect caused by moving strings. It produces line-like discontinuities on the CMB temperature anisotropies ~\citep{Hindmarsh:1993pu,Stebbins:1995} of the form
\begin{equation}
    \frac{\delta T}{T} \sim 8\pi G\mu v_{\uS}.
\end{equation}
Here $v_\uS$ is the transverse velocity of the string. 
The CS network is also expected to produce extra CMB polarization \citep{Benabed:1999wn,Danos:2010gx,Brandenberger:2011eq,Bevis:2007qz} and dipole modulation  \citep{Ringeval:2015ywa}.
 
CMB-based approaches to search for CS are quite diverse. For example, \cite{Ade:2013xla,Ade:2015xua} use  the {\it Planck} temperature power spectrum to get an upper bound of $G\mu<3.0\times10^{-7}$ for Abelian-Higgs strings, which improves with the {\it Planck} polarization\footnote{Note that the {\it Planck} 2015 polarization data is preliminary
at large scales due to residual systematics.} to $G\mu<2.0 \times 10^{-7}$ for Abelian-Higgs strings \citep{Lizarraga:2016onn}, to $G \mu < 1.5 \times 10^{-7}$ for Nambu-Goto strings \citep{Lazanu:2014eya} and to $G\mu<1.1 \times 10^{-7}$ for a multi-parameter fit to the unconnected segment model \citep{Charnock:2016nzm}.
In the search for the CS network, one could exploit the non-Gaussianity of CS-induced fluctuations, e.g., through measuring CMB bispectrum, using Wavelet-based methods, or measurements of the CMB Minkowski functionals. These searches  lead to 
$G\mu<8.8\times10^{-7}$,
$G\mu<7\times10^{-7}$ and $G\mu<7.8\times10^{-7}$, respectively
\citep{Hindmarsh:2009qk, Hindmarsh:2009es, Ade:2013xla,Regan:2015cfa, Ringeval:2010ca, Ducout:2012it}. 
%
%

Examples of real-space-based statistical methods are using the crossing statistics  of CMB maps which yields the detectability level of $G\mu\gtrsim 4.0\times 10^{-9}$ for noise-free simulations \citep{Movahed:2010zq}, and using  the unweighted Two-Point Correlation Function (TPCF) of CMB
peaks which gives $G\mu\gtrsim 1.2\times 10^{-8}$
 for noiseless, 1'-resolution maps \citep{Movahed:2012zt}. 
 Some methods exploit the specific KS pattern, i.e, the line-like discontinuities of CMB fluctuations. 
\cite{Stewart:2008zq} applied edge-detection
algorithms to find a minimum detectability of $G\mu \gtrsim
5.5\times 10^{-8}$ for a South Pole Telescope-like scenario and \cite{Hergt:2016xup} used wavelet
and curvelet methods to claim a detection level of  
$G\mu\gtrsim 1.4 \times 10^{-7}$ for the third generation SPT.

In a recent paper we introduced a pipeline that applied various image processing and statistical tools to investigate the detectability of the CS network imprint on CMB temperature maps \citep{Sadr:2017hfm}. We claimed CS detectability for strings with $G\mu\gtrsim 4.3\times 10^{-10}$ for noiseless, $0.9'$-resolution, $7.2^\circ \times 7.2^\circ $ patches , and 
  with $G\mu\gtrsim 1.2 \times 10^{-7}$ for CMB-S4-like (II) experiments.
There are also the quite recent neural network-based approaches, giving $G\mu \gtrsim 2.3 \times 10^{-9}$ for noiseless arcminute-resolution maps \citep{Ciuca:2017wrk}. \cite{Ciuca:2017gca} use a convolutional neural network to locate the position of the CSs and get a limit of $G\mu \gtrsim 5 \times 10^{-9}$. 
 The tightest bound on the CS tension, $10^{-14} \le G\mu \le1.5 \times 10^{-10}$, comes from the gravitational wave emission of Nambu-Goto CS loops ~\citep{Ringeval:2017eww, Blanco-Pillado:2017oxo,Blanco-Pillado:2017rnf}. It should be noted, however, that these constraints strongly depend on the string microstructure. Abelian-Higgs field-theory simulations indicate that string loops decay mainly by the emission of massive radiation and emit less gravitational waves than estimated from Nambu-Goto simulations \citep{Hindmarsh:2017qff}, thus weaken the bounds.
Constraints from CMB maps are therefore more robust and conservative.

In this work we propose to use machine learning (ML)-based algorithms to search for the KS imprint of the CS network on CMB data.  The goal is to develop a detection strategy capable of putting  the tightest upper bound on the CS tension through optimally exploiting the available information accessible to the multi-scale pipeline of \cite{Sadr:2017hfm}. For this purpose, we choose to use two tree-based supervised classifiers: random forest (RF) and gradient boosting (GB).

In the following, after introducing our simulations (Section~\ref{simulation}), we explain our proposed strategy for CS detection from CMB maps,  through compressing the map information into feature vectors (Section~\ref{Pre-proc}). The vectors are passed to tree-based ML methods to search for CS imprints (Section~\ref{Post-proc}). 
We then describe in detail our proposed strategy in reporting the results in cases with possibly biased measurements (Section ~\ref{stdetection}). 
Finally we present the results (Section~\ref{results}) and conclude with the discussion of the results (Section~\ref{sum_conc}).
\section{SIMULATIONS}\label{simulation}
Our simulations of the CMB sky closely follow \cite{Sadr:2017hfm} and consist of three components: the Gaussian contribution $G$ (including the primordial inflationary fluctuations, as well as the secondary lensing effect), CS-induced perturbations given by $G\mu \times S$, with $S$  describing the simulated normalized template for the CS signal  ~\citep[using the Bennett-Bouchet-Ringeval
code,][]{Bennett:1990, Ringeval:2005kr} and $G\mu$ setting its overall amplitude, and the experimental noise $N$, described by white Gaussian random fields parametrized by the corresponding $SNR$ (signal to noise ratio). Our 2-Dimensional sky map $T(x,y)$ is thus given by:
\be\label{fullmap}
T(x,y)=B[G(x,y)+G\mu \times S(x,y)]+N(x,y).
\ee 
 $B$ denotes the beam function, here taken to be the model used in some ground-based observations \citep{Fraisse:2007nu,White:1997wq}, with an effective FWHM$\approx0.9'$, as well as a Plank-like Gaussian beam with  FWHM$\approx5'$.
The simulated maps are square patches with sides $\Theta=7.2^\circ$, pixelized  into squares with resolution $R=0.42'$. This yields a total of $1024\times 1024$ pixels. For more details see \cite{Sadr:2017hfm}. 
\section{DETECTION STRATEGY I: Pre-processing}  \label{Pre-proc} 
 \label{detect_strategy}
The CS detection algorithm of this work has two main steps. The pre-processing step compresses information from maps into  feature vectors (each with $275 $ elements). The feature vectors are then passed to the classifier unit for classification.  These two steps are briefly explained in this section and the  following. 

\begin{figure}
\begin{center}
\includegraphics[scale=0.32]{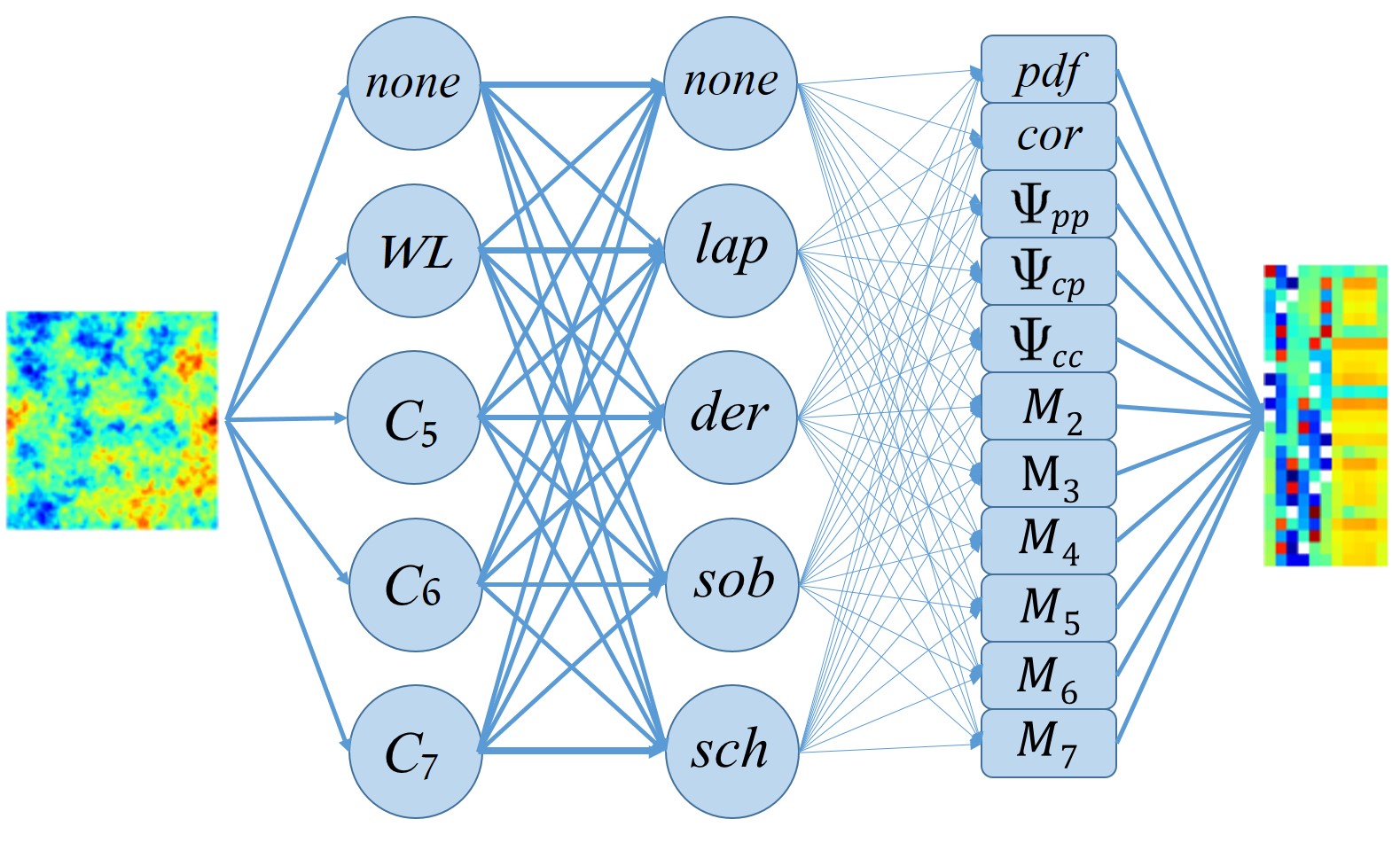}
\end{center}
\caption{A schematic view of the feature vector generation. For a CMB map (input on left side) it produces a $275$-dimensional feature vector, here presented as a $25\times11$ array (right side). The vector includes all possible combinations of decomposers, filters and statistical measures used in this work.\label{fig:pre}}
\end{figure}

\begin{figure}
\begin{center}
\includegraphics[scale=0.18]{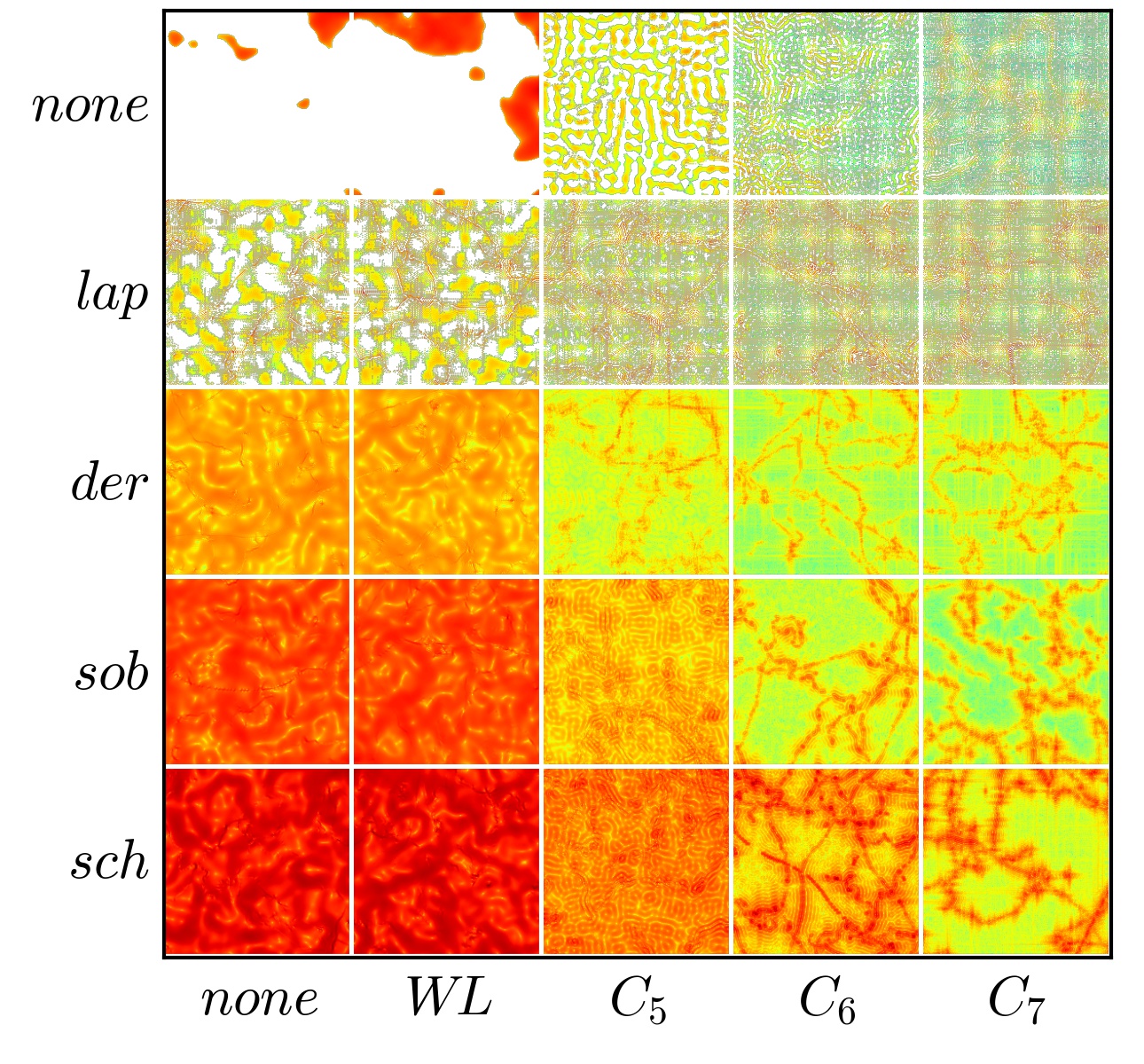}
\end{center}
\caption{All of the $25$ outputs of the image processing layers of the algorithm applied to a map with $G\mu=1.0\times10^{-7}$. The color scale is logarithmic. These are then passed to the 11 statistical measures, yielding the full set of 275 features.
}
\label{fig:image_processing}
\end{figure}

The feature extraction step employs three layers of image processors and statistical measures to produce a feature vector as the input for the learning unit (Figure~\ref{fig:pre}).
The first two layers aim at producing maps with enhanced CS detectability (Figure~\ref{fig:image_processing}), and the third layer quantifies the deviation of certain statistical measures of the map from those of the baseline model corresponding to null simulations with no CS imprints. These layers can be briefly described as: \\
(i) decomposers to disintegrate maps into scales relevant to the signal of interest. 
The output is  labeled as either {\it none} (corresponding to the full map), {\it WL} (or wavelet\footnote{The wavelet used here is the Daubechies db12 \citep{daubechies1990wavelet} with the mother function provided by the {\it PyWavelets} package, \url{https://github.com/PyWavelets}, and with the coefficients low-pass filtered with a threshold of $3$.}), 
%
or one of the three curvelet components $C_5$, $C_6$ and $C_7$, corresponding to the three smallest scales\footnote{We used the {\it Pycurvelet} package \citep{Sadr:2017hfm} as our 2D, discrete version of the curvelet transform \citep{candes2006fast}. This package is the python-wrapped version of {\it CurveLab}, \url{http://www.curvelet.org/}.  We chose $n_{\rm scales}=7$ and $n_{\rm angles}=10$ as the curvelet transformation parameters.}\citep{Sadr:2017hfm}.\\
(ii) various filters to enhance edges. 
The output is  labeled as either {\it none} (corresponding to the full map),  {\it der} (or derivative),  {\it lap} (or Laplacian), {\it sob} (or Sobel) or {\it sch} (or Scharr). \\
(iii) different statistical measures applied on the filtered, scale-decomposed maps. The measures are {\it pdf} (the probability distribution function), $M_2$ to $M_7$ (the second to seventh statistical moments), {\it cor} (the map correlation function),  $\Psi_{pp}$ (the autocorrelation of peaks),  $\Psi_{cc}$ (the autocorrelation of upcrossings) and  $\Psi_{cp}$ (the peak-upcrossing cross-correlation).
For a thorough description see \cite{Sadr:2017hfm}. See also \citealt{rice1944mathematical,Bardeen:1985tr,Bond:1987ub,Ryden:1988rk,ryden1988collapse,Landy:1993yu,Matsubara:1995wj,Matsubara:2003yt,Ducout:2012it,Pogosyan:2008jb,Gay:2011wz,Codis:2013exa}.
For any given map, the final output of the pre-processor is a feature vector with $275$ elements, corresponding to all combinations of processors from each layer (Figure~\ref{fig:pre}). 
The feature vector is then passed to the learning unit for classification, i.e. to RF and GB,  to learn from simulations and to estimate $G\mu$ for new maps.


\section{DETECTION STRATEGY II: Learning Process}  \label{Post-proc} 

In this section we develop a machine-based algorithm to estimate the $G\mu$'s of given CMB maps using their feature vectors generated by the pre-processors. 
%
We use supervised classifiers to build the data-driven model which maps the feature vector $\bf{x}$ to the predictor $\bf{y}$.
More specifically, we use the two powerful decision-tree-based ensemble methods:  random forest or RF \citep{breiman2001random} and gradient boosting or GB \citep{friedman2001greedy} that combine a set of weak learners to improve the prediction performance \citep{quinlan1986induction,kearns1988thoughts,opitz1999popular,polikar2006ensemble,rokach2010ensemble}. 
The tree, with its top-down greedy structure, starts from its root corresponding to the full set of observations and splits successfully into branches producing the prediction space. The branching process is based on dividing samples into homogeneous sets considering the most significant differentiator in input variables. 

The RF classifier is based on growing many decision trees and its prediction will be the decision with the highest vote from all trees. The GB classifier, on the other hand, is based on the gradual improvement of a sequence of models toward better prediction, usually with decision trees as their base learners. 
This is achieved through improving the model in a stage-wise manner by reducing an arbitrary loss function, here taken to be $\mathcal{L}=\sqrt{\ev{(\bf{y}-{\bf y}_{\rm fid})^2}}$
where $\bf{y}$ and ${\bf y}_{\rm fid}$ are the model and true (fiducial) values, respectively \citep{friedman2001greedy}.

It is important to note that not all features are expected to be independent or equally significant. 
The tree-based learners report the importance of each feature based on the impact of its changes on the classifying parameter, here $G\mu$. This is called feature importance analysis and we will use it to find elements of the feature vectors with the most significant roles in CS detection. 
Feature analysis can help to enormously reduce the dimension of the feature space 
without a practical impact on the machine's performance \citep{bermingham2015application}. 
Extra care needs to be taken in dimensionality reduction  since a too small number of features may lead to experiment-dependent models with little generality. 
In this work, we investigate the importance of features by averaging their number of occurrences among the top ten features through all machine learning models (MLMs). 


Overfitting is a common problem in  non-parametric algorithms due to their extreme flexibility. 
In overfitting, noise and random correlations of the training set impact the model and result in reduced sensitivity when confronted with new observations that do not have those spurious features.
Cross-validation or CV, also used in this work, is a common powerful technique to avoid this issue \citep{kohavi1995study}.
The training set is partitioned into smaller training sets as well as a validation set. The model is made using the former while the latter plays the role of a new observation to assess how smoothly the method generalizes to new datasets.
Here we use a $K$-fold CV strategy where the original dataset is randomly divided into $K$ equal subsets with $K-1$ subsets forming the training sets and one the validation set. The process is repeated $K$ times to guarantee each subset is validated once. 

We divide the $G\mu$ range used in this work ($2.5\times 10^{-11}<G\mu<5\times10^{-7}$ ) into $N_{\rm class}=18$ classes, with equal separation in $\ln G\mu$. A null class with $G\mu=0$ is also considered.  
The machine is trained by applying the RF and GB algorithms  
as MLMs to the feature vectors of $N_{\rm train}=1900$ CMB maps, corresponding to $N_{\rm sim}=100$ simulations for each class.  
Our training unit has $N_{\rm L}=100$ MLMs with different seeds, each with a $K=10$-fold cross validation.  
In each folding 90 maps are used for training and the remaining 10 maps of the class are used as the test set.
 The results have been tested for robustness against various foldings. 
To get a better control of the overfitting problem, we also generate a separate validation set with ten maps for each class.
This MLM can then be applied to any given CMB map to estimate its level of CS contribution. 

The trained MLM assigns to any input map a probability vector $\vec{P}$, corresponding to the $G\mu$  of the classes ($\vec{G\mu}$). 
We report the following (Bayesian) weighted average of the $G\mu$ as the {\it predicted} $G\mu$:
\begin{equation}\label{gmu_obs}
G\mu_{\rm pre}=\vec{P}.{\vec{G\mu}}.
\end{equation}
Classifiers often suffer from the limitation that the classes do not necessarily include the underlying parameter of an observation.
The above weighted averaging  partially alleviates this problem. It should be noted that a relatively flat $\vec{P}$ would reflect the limited power of the trained MLM in discriminating between classes.



In the next section, we clarify in detail how we report the machine's output and translate it to the language of CS detection  and measuring its contribution.

\section{Detecting strings or measuring their tension?}\label{stdetection}
Applying the detection strategy of the previous section yields a distribution of  $G\mu_{\rm pre}$ for any of the $G\mu_{\rm fid}$ classes.
This distribution is ideally peaked around the $G\mu_{\rm fid}$, and its dispersion is sourced by cosmic variance, as well as  contamination from primordial anisotropies and experimental noise. There is also a subdominant contribution to the fluctuations of $G\mu_{\rm pre}$ caused by the random seed of the MLMs which would decrease as the number of MLMs increases. 

We define the minimum {\it detectable} $G\mu$, or $G\mu_{\rm det}$, as the minimum $G\mu$ whose distribution can be distinguished, with a maximum two-tail P-value of $0.0054$, from all other $G\mu$ classes, including the null class. 
This minimum detection states there is a significant deviation in the map from the null hypothesis (with no string input). Note that this does not necessarily imply an unbiased measurement of the CS tension.  
We therefore define the minimum {\it measurable} $G\mu$, or $G\mu_{\rm mes}$, as the minimum $G\mu$ above which the $G\mu_{\rm pre}$'s are unbiased. More precisely, $G\mu_{\rm mes}$ is the minimum $G\mu$ whose bias, defined as $G\mu_{\rm bf}-G\mu_{\rm fid}$ is smaller than one sigma. Here $G\mu_{\rm bf}$ is the best-fit $G\mu$ in the distribution of $G\mu_{\rm pre}$.

The next section presents the results of applying the proposed strategy to simulated CMB maps corresponding to several experimental cases  \citep{Sadr:2017hfm}.

\section{RESULTS}\label{results} 
\begin{figure*}
\begin{center}
\includegraphics[scale=0.55]{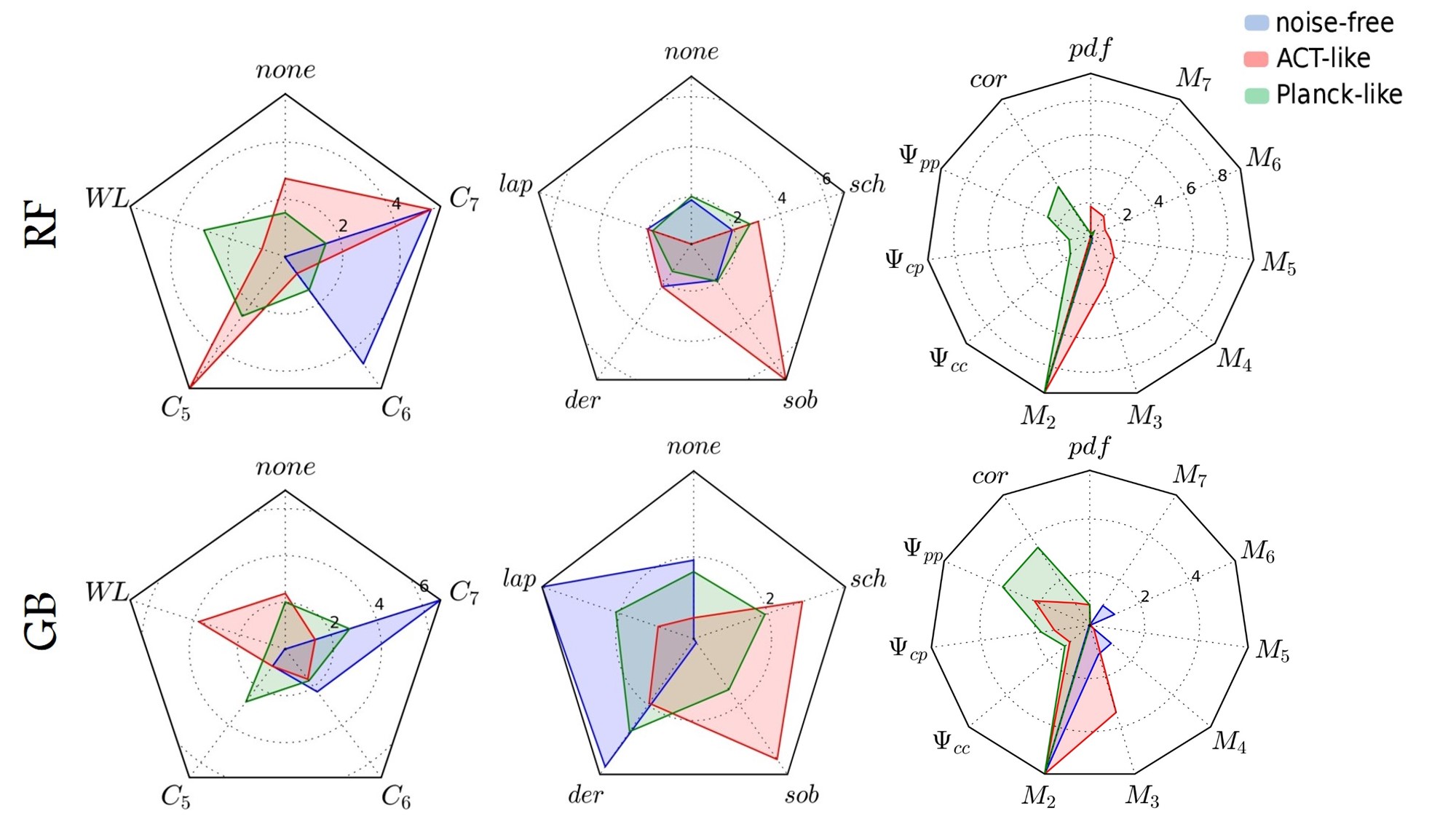}
\end{center}
\caption{Feature importance report: the average number of times each feature appeared among the top ten features, for each layer of the pre-processor, for the RF (top) and GB (bottom) learner. }
\label{features}
\end{figure*}

In this work, we simulate CMB maps for five experimental setups: an ideal noise-free case, two CMB S4-like experiments, an ACT-like and a Planck-like case. The Planck-like simulations  are smoothed with a Gaussian beam with $FWHM=5'$, while for the other four cases the effective beam is $FWHM=0.9'$.
The details of the experimental settings are given in \cite{Sadr:2017hfm}.
The feature vector for each map is generated through its pre-processing, which is then passed to the tree-based learning unit of the algorithm.   
 Figure~\ref{features} compares the feature importance of the three pre-processing layers for noise-free, ACT-like and Planck-like cases, and for the two tree-based algorithms considered in this work, namely RF and GB.
We find that the sixth and seventh curvelet components of the input maps have the dominant role in the first pre-processing layer for the noise-free case. That is expected since these components contain the small-scale information which is important for CS detection. 
On the other hand, the instrumental noise contaminates the small scales most, making part of the CS signal  in these higher modes inaccessible. That explains the more important roles of $C_5$ and $WL$
for ACT-like and Planck-like setups. 
The middle panels of Figure~\ref{features} indicate that the classifiers have no significant preference for the filters. However, the ACT-like scenario should be excepted where Sobel seems to have a major impact on the results if the RF classifier is used.
In the third layer, the second moment of the filtered maps is clearly the main player in both RF and GB algorithms (right panels of Figure~\ref{features}). 
The results from feature analysis could enormously decrease the computational cost of future analysis by  helping to limit the training process to the feature subspace with most significant impact on the classification.

Table~\ref{tabledet} presents the predictions of our proposed tree-based detection strategy for the minimum detectable $G\mu_{\rm det}$.
\begin{table}
\centering
\caption{The minimum detectable $G\mu$, or $G\mu_{\rm det}$, for the two tree-based algorithms, GB and RF, and for the five experimental setups.}
\label{tabledet}
\begin{tabular}{ccc}
   experiment   &     $ G\mu_{\rm det}({\rm GB}) ~~~$  & $ G\mu_{\rm det} ({\rm RF}) $ \\ 
\hline

  noise-free   &    $4.3 \times 10^{-10}$ &  $2.1 \times 10^{-10}$ \\
  CMB-S4-like (II)   &    $1.2 \times 10^{-7}$ &   $3.0 \times 10^{-8}$ \\
  CMB-S4-like (I)    &   $1.2 \times 10^{-7}$ &   $1.2 \times 10^{-7}$ \\
   ACT-like              &    $1.2 \times 10^{-7}$ &   $1.2 \times 10^{-7}$ \\
   Planck-like                 &     $7.0 \times 10^{-7}$ &   $5.0 \times 10^{-7}$ \\
 \hline
\end{tabular}
\end{table}
Similarly, Table~\ref{tablemes} reports the predicted $G\mu_{\rm mes}$'s for the various experiments considered in this work. 
We find that for a Planck-like experiment the detection limit is $G\mu_{\rm det} \approx 5\times 10^{-7}$, while the $G\mu_{\rm mes}$ is above the upper bound of $G\mu$ range considered in the simulations of the training process. This means that our method is capable of detecting traces of CS with high significance for a $G\mu$ as low as $5\times 10^{-7}$. However, this method, with its current $G\mu$ range and fiducial classes, can not  make an unbiased measurement of such small string tensions.
For a noise-free observation of the sky, the algorithm can distinguish the traces of  CS networks down to $2.1 \times 10^{-10}$, and can correctly estimate the level of CS contribution for $G\mu$ above $G\mu_{\rm mes} \approx 3.6 \times 10^{-9}$. 

Note that Table~\ref{tablemes} only reports the minimum measurable $G\mu$'s and not their associated errors. That is because the uncertainties in our measurements are dominated by the bin size of $G\mu$ classes, and not the statistical error. Therefore, for a class with $G\mu_{\rm fid}$, the uncertainty in the measurement is $\sigma_{G\mu}=\Delta \ln G\mu \times G\mu_{\rm fid}$, irrespective of the experiment.

\begin{table}
\centering
\caption{Similar to Table~\ref{tabledet} but for minimum measurable $G\mu$'s, or $G\mu_{\rm mes}$'s.}
\label{tablemes}
\begin{tabular}{ccc}
   experiment   &     $ G\mu_{\rm mes}({\rm GB})~~~ $  & $ G\mu_{\rm mes} ({\rm RF}) $ \\ 
\hline
    noise-free &  $3.6 \times 10^{-9}$ &  $3.6 \times 10^{-9}$ \\
    CMB-S4-like (II)  &  $1.2 \times 10^{-7}$ &  $1.2 \times 10^{-7}$ \\
    CMB-S4-like (I)  &  $1.2 \times 10^{-7}$ &  $2.5 \times 10^{-7}$ \\
     ACT-like  &  $2.5 \times 10^{-7}$ &  $2.5 \times 10^{-7}$ \\
     Planck-like &    $1.0 \times 10^{-6}$ &             $1.0 \times 10^{-6}$ \\

\hline
\end{tabular}
\end{table}

\section{DISCUSSION}\label{sum_conc}

We proposed a tree-based machine learning algorithm for detecting and measuring the trace of CS-induced signals on CMB maps, simulated for various observational scenarios. Our simulations consisted of $1900$ maps, passed through the pre-processing unit of the algorithm to form the feature vectors, which are the inputs to the classifiers. 
The simulations correspond to $18$ classes of $G\mu$ in the range $G\mu = 2.5\times 10^{-11}$ to $5\times 10^{-7}$, with equal spacing in $\ln G\mu$, and one null class.
Out of these maps, $90\%$ were used for training the classifiers (here taken to be random forest and gradient boosting) and the rest as test sets.  
We performed feature  analysis on the feature vectors to find the significance of the role of each feature for the classification. The results can be a major help in reducing the computational cost of future analysis by decreasing the dimension of the feature space and limiting the analysis  to the most significant features. 
As general results we can state that the scale of curvelet components should be matched to the effective resolution of experiments in the presence of experimental noise, larger-scale curvelet components are the more important decomposers. For filters it is difficult to make a definite recommendation, 
 while the second moment is the most important statistical measure in the classification process. 

We find that, for each experimental case, three $G\mu$ regimes can be distinguished, whose boundaries marked by the $G\mu_{\rm det}$ and $G\mu_{\rm mes}$. For $G\mu$'s greater than $G\mu_{\rm mes}$, the algorithm is capable of measuring the CS contribution, with no bias and with an error determined by the bin size of that class. For $G\mu$'s smaller than $G\mu_{\rm mes}$ but larger than $G\mu_{\rm det}$, the algorithm can detect the signal, but cannot always make an unbiased measurement of its level. 
For $G\mu$'s smaller than $G\mu_{\rm det}$, the CS signals are not reliably detected by the algorithm.
The predicted $G\mu_{\rm det}$ for a noise-free experiment is $2.1 \times 10^{-10}$.
This bound is, to the best of our knowledge, well below 
the claimed detectability levels by other methods on noise-less maps. Compare, e.g., to the detectability bound of $G\mu\gtrsim 4.0\times 10^{-9}$ from crossing statistics \citep{Movahed:2010zq}, $G\mu\gtrsim 1.2\times 10^{-8}$ from the unweighted Two-Point Correlation Function of CMB peaks  \citep{Movahed:2012zt}, $G\mu \gtrsim 6.3 \times 10^{-10}$  from Wavelet domain Bayesian denoising algorithm \citep{Hammond:2008fg}, $G\mu \gtrsim 2.3 \times 10^{-9}$ from the Neural network-based approaches \citep{Ciuca:2017wrk} and $G\mu\gtrsim 4.3\times 10^{-10}$  from a multi-scale pipeline for CS detection \citep{Sadr:2017hfm}.  The minimum detectable tension in this work for a CMB-S4-like (II) experiment, $G\mu\gtrsim 3.0 \times 10^{-8}$, is a major improvement over the claimed detectability level by the above multi-scale pipeline, $G\mu\gtrsim 1.2 \times 10^{-7}$.
 %
    For a Planck-like case, the minimum detectable $G\mu$ is $5\times 10^{-7}$, comparable to the current upper bounds from Planck data \citep{Ade:2013xla}. Both classification methods seem to perform at a similar level, with RF appearing slightly more powerful based on the numbers in Tables \ref{tabledet} and \ref{tablemes}.

An important and immediate improvement to this work is to devise and apply debiasing techniques to remove the gap between $G\mu_{\rm mes}$ and $G\mu_{\rm det}$. 
 Given the continuous nature of the problem, one might also expect that using a regressor would improve the results.
 That is because classifiers, by construction, are the method of choice in categorization problems while  regressors in general  are more suited for parameter estimation with continuous parameter ranges. It should be noted that using Bayesian averaging (Eq. \ref{gmu_obs}) in the parameter measurement step partially converted the classifiers of this work into regressors. We leave the full treatment of regressor-based algorithms for CS detection to future work.
 

\section*{Acknowledgements}
The authors would like to thank C. Ringeval, F. R. Bouchet  and the Planck collaboration as they generously provided the simulations of the cosmic string maps used in this work. A. Vafaei acknowledges helpful discussions with  Y. Fantaye. The numerical simulations were carried out on Baobab at the computing cluster of University of Geneva.  





\bibliography{bib}{}
\bibliographystyle{mnras}

\bsp	
\label{lastpage}
\end{document}